\documentclass[twocolumn,aps,prl,superscriptaddress, amsmath,amssymb,longbibliography]{revtex4-1}

\usepackage{graphicx}
\usepackage{dcolumn}
\usepackage{bm}
\usepackage{epstopdf}
\usepackage{xcolor}
\usepackage{siunitx} 

\newcommand{\AS}[1]{{\color{black} #1}}
\newcommand{\Bo}[1]{{\color{black} #1}}

\newcommand{\Anton}[1]{{\color{black} #1}}

\begin{document}

\title{Oscillatory chiral flows  in confined active fluids with obstacles }

\author{Bo Zhang}
\affiliation{Materials Science Division, Argonne National Laboratory, 9700 South Cass Avenue, Lemont, IL 60439, USA}

\author{Benjamin Hilton}
\affiliation{Department of Physics, University of Bath, Claverton Down, Bath BA2 7AY, UK}

\author{Christopher Short}
\affiliation{Department of Physics, University of Bath, Claverton Down, Bath BA2 7AY, UK}

\author{Anton Souslov}
\email{a.souslov@bath.ac.uk}
\affiliation{Department of Physics, University of Bath, Claverton Down, Bath BA2 7AY, UK}

\author{Alexey Snezhko}
\email{snezhko@anl.gov}
\affiliation{Materials Science Division, Argonne National Laboratory, 9700 South Cass Avenue, Lemont, IL 60439, USA}

\date{\today}
\begin{abstract}
An active colloidal fluid comprised of self-propelled spinning particles injecting energy and angular momentum at the microscale demonstrates spontaneous collective states that range from flocks to coherent vortices. Despite their  seeming simplicity,  the emergent far-from-equilibrium behavior of these fluids remains poorly understood, presenting a challenge to the design and control of next-generation active materials. When confined in a ring, such so-called polar active fluids acquire chirality once the spontaneous flow chooses a direction. In a perfect ring, this chirality is indefinitely long-lived. Here, we combine experiments on self-propelled colloidal Quincke rollers and mesoscopic simulations of continuum Toner-Tu equations to explore how such chiral states can be controlled and manipulated  by obstacles. For different obstacle geometries three dynamic steady states have been realized: long-lived chiral flow, an apolar state in which the flow breaks up into counter-rotating vortices and an unconventional collective state with flow having an oscillating chirality. The chirality reversal proceeds through the formation of intermittent vortex chains in the vicinity of an obstacle. We demonstrate that the frequency of collective states with oscillating chirality can be tuned by obstacle parameters. We vary obstacle shapes to design chiral states that are independent of initial conditions. Building on our findings, we realize a system with two triangular obstacles that force the active fluid towards a state with a density imbalance of active particles across the ring. Our results demonstrate how spontaneous polar active flows in combination with size and geometry of scatterers can be used to control dynamic patterns of polar active liquids for materials design.

\end{abstract}

\maketitle


\section{I. Introduction}

Active matter is a class of far-from-equilibrium systems whose components transduce energy from the environment into mechanical motion~\cite{marchetti2013hydrodynamics, bechinger2016active, martin2013driving}.
One remarkable feature of active-matter systems is their natural tendency towards spontaneous formation of large-scale collective behavior and dynamic self-assembled architectures~\cite{sokolov2007concentration, vicsek2012collective, snezhko2009self, zottl2016emergent, sanchez2012spontaneous, snezhko2011magnetic, yan2016reconfiguring,demortiere2014self}.
Recent advances in the design of active colloids have enabled unprecedented control of flow in a class of active systems in which rolling couples rotation to propulsion~\cite{bricard2013emergence, kaiser2017flocking, driscoll2017unstable, kokot2017active, soni2019odd, martinez2018emergent, kokot2018manipulation}.

The emergent dynamic patterns and their corresponding transport phenomena have been widely studied in different living or synthetic systems such as bacterial suspensions, mixtures of molecular motors and filamentary proteins, Janus particles, as well as field-driven rollers~\cite{ballerini2008interaction, cavagna2014bird, dombrowski2004self, sokolov2012physical,  decamp2015orientational, wu2017transition, huber2018emergence,  ross2019controlling, narayan2007long, kudrolli2008swarming,  yan2016reconfiguring, driscoll2017unstable, rubenstein2014programmable, han2020emergence, scholz2018rotating}.
 Complex and crowded environments can often lead to exotic collective response of active systems~\cite{bechinger2016active}.\AS{For instance, a suspension of swimmers under channel confinement form a steady unidirectional circulation~\cite{wioland2016directed}, but when   confined in a two-dimensional circular chamber self-organizes into a single stable vortex~\cite{wioland2013confinement}, whereas hydrodynamically coupled chambers with swimmers give rise to a self-assembled vortex lattice with the vorticity at each lattice site exhibiting either ferromagnetic or antiferromagnetic order~\cite{wioland2016ferromagnetic, beppu2017geometry}. Periodic arrays of microscopic pillars introduced into bacterial suspensions have promoted the formation of emergent self-organized arrays of alternating vortices controlled by pillar spacing, with the order manipulated using chiral pillars~\cite{nishiguchi2018engineering}. Multitude of hydrodynamic instabilities has been reported in active nematic suspensions under confinement~\cite{voituriez2005spontaneous, marenduzzo2007steady, giomi2012banding, shendruk2017dancing,coelho2019active,fielding2011nonlinear}.  } Furthermore, the design of the environment can be used to capture and concentrate active particles~\cite{kaiser2012capture, galajda2007wall, koumakis2013targeted}, to direct transport using ratchet effects~\cite{reimann2002brownian, di2010bacterial, sokolov2010swimming},
or to couple to state variables such as pressure~\cite{solon2015pressure}.

Past studies have shown that obstacles can disrupt coherence and collective response in active fluids~\cite{morin2017distortion,kokot2018manipulation}. Here, we instead use obstacles as a tool for materials design by controlling emergent collective chiral states.
We combine experiments and simulations to reveal that interactions of the chiral fluid with a passive scatterer can promote reversal of the flow away from a global chiral state. These observations let us design and tune exotic collective states with oscillating chiral flows. The chirality reversal process proceeds through an intermittent formation of vortex chains in the vicinity of the obstacle. We show that oscillation dynamics of the chiral flows and also the density distribution of active particles could be effectively controlled through scatterer geometry.


\begin{figure*}[t]
\centering
\includegraphics[width=1.0\linewidth]{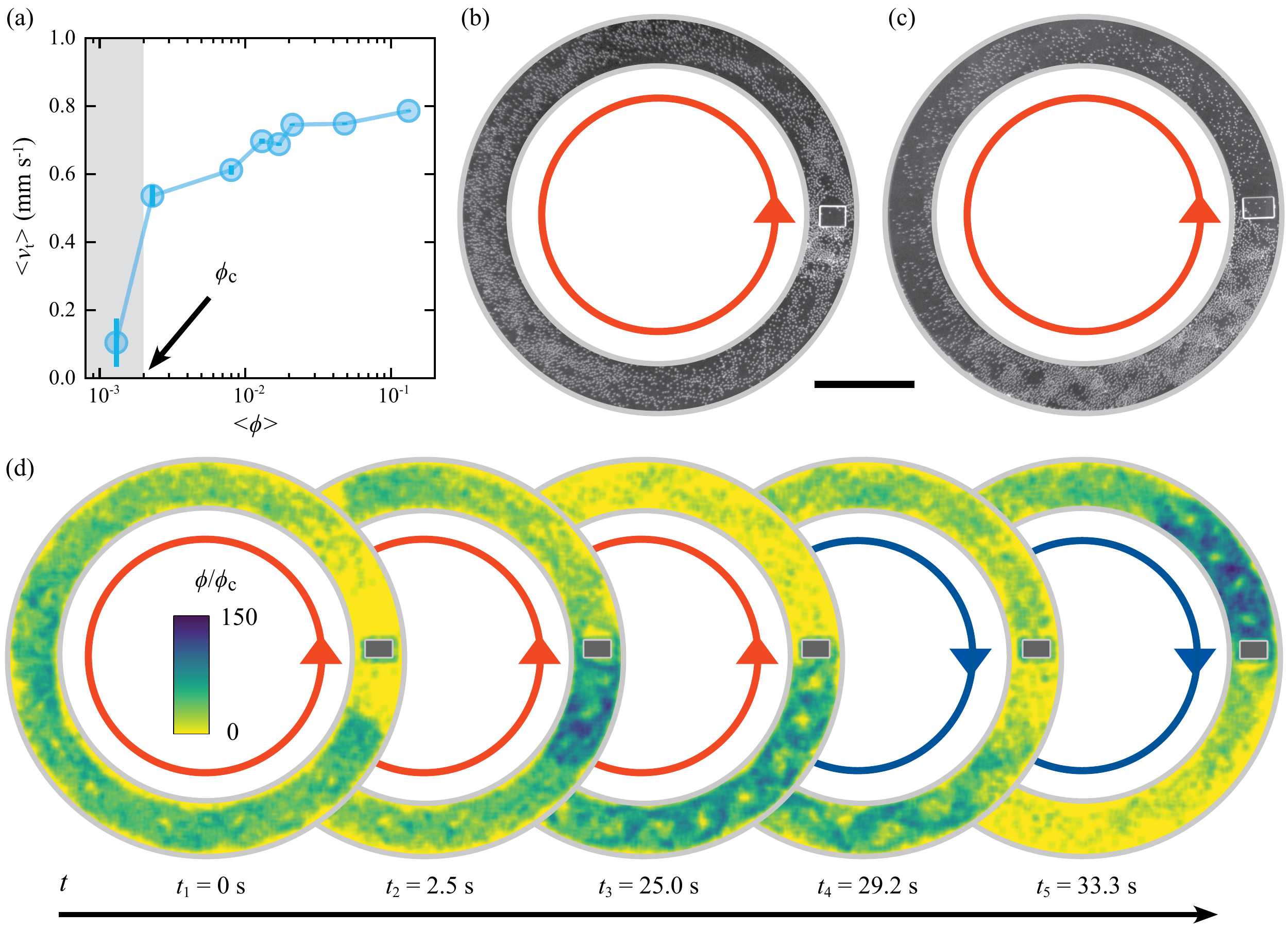}
\caption{
Obstacle-induced chirality reversal.
	(a) Average tangential velocity of rollers at different area fractions in a track. The width of the track is given by $W$ = 0.25 mm. Error bars indicate the standard error of velocities. The critical area fraction is given by $\phi_\text{c}$ = 0.002, above which polar bands (or, more generally, a polar active fluid state) are formed in the track.
	(b) Polar active fluid composed of Quincke rollers in a track with an obstacle. The width $W$ of the track is 0.25 mm. The width of the obstacle $W_\text{ob}$ = 0.5 $W$. Rollers are scattered near the incoming side of the obstacle. The red circle with an arrow represents the CCW flow direction. \Bo{The initial chirality of the flow (CCW or CW) is spontaneously selected by the system and it is different from experiment to experiment}
	(c) Same as (b), but with a wider obstacle, $W_\text{ob}$ = 0.6 $W$. More rollers are scattered by the obstacle and form intermittent patterns. The active fluid eventually reverses the flow direction from CCW to CW. (See Supplementary Movie 1).
	(d) Typical time evolution of the distribution of local area fraction of the rollers during one chirality reversal event. Rollers first form a polar band ($t_1$). Once the polar band arrives at the obstacle, rollers are scattered ($t_2$), and then dynamically assemble into the intermittent pattern with high densities($t_3$). Then, the active fluid reverses the flow direction ($t_4$) and  rollers are now scattered on the other side of the obstacle ($t_5$).
	Blue and red circles with an arrow shows the CW and CCW flows of particles, respectively (see also Supplementary Video 1).
	In (a-b, d), the scale bar is 0.5 mm; the electric field strength is $E=$ 2.7 V \SI{}{\micro\meter}$^{-1}$; the area fraction is $\langle \phi \rangle =45 \phi_\text{c}$.
}
\label{Fig1}
\end{figure*}

\begin{figure*}[t]
\centering
\includegraphics[width=1.0\linewidth]{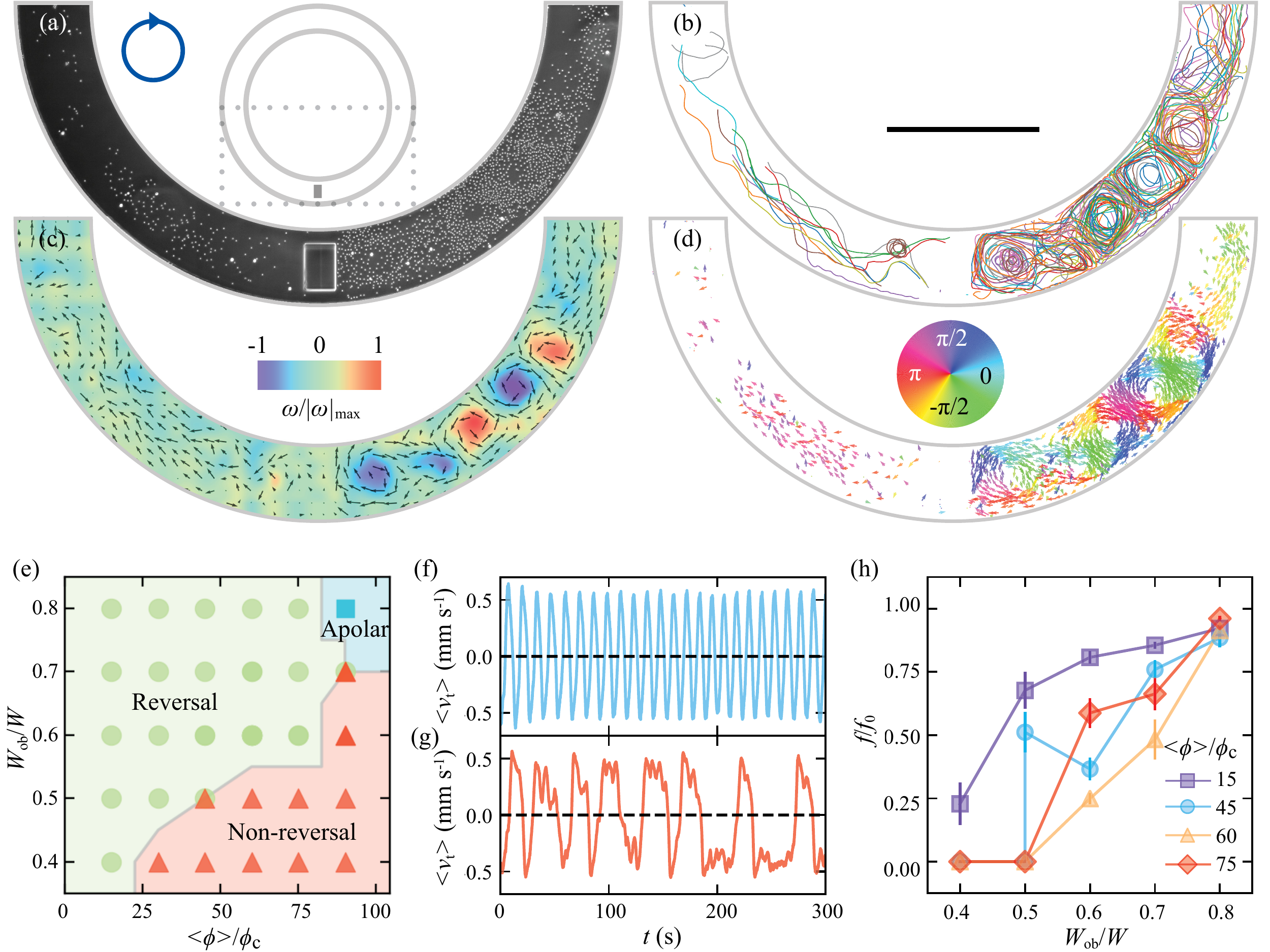}
\caption{
Chirality reversal in a track with a rectangular obstacle.
	(a) Microscopic images of a polar active fluid. Multiple vortices form near the obstacle. The polar order is partially restored after rollers pass \Bo{around} the obstacle. (Flow is CW; only lower halves of the tracks are shown, see also Supplementary Movie 2).
	(b) Particle trajectories of rollers. Some rollers inside each vortex follow square trajectories due to the effect of confinement. The duration of particle trajectories is 1.2 s. Only 5\% of particle trajectories are shown.
	(c) Superimposed velocity (arrows) and vorticity (background color) fields of the rollers. See also Supplementary Movie 2.
	(d) Individual particle velocity. For a better visualization, the velocity direction is colored according to the color wheel. See also Supplementary Movie 2.
	In (a-d), $W_\text{ob}$ = 0.6 $W$; $\langle \phi \rangle=15 \phi_\text{c}$; the scale bar is 0.5 mm.
	(e) Phase diagram of chirality reversal and non-reversal for different obstacle widths $W_\text{ob}/W$ and average area fractions $\langle \phi \rangle$. Shown are the chirality-reversal phase (green circles), the phase with no chiraliry reversal (red triangles) and the apolar vortex phase (blue squares). Coexistence is indicated by superimposing different symbols.
	(f-g) Time evolution of average tangential velocity of fast (f) and slow (g) reversals with values of $W_\text{ob}/W$ = 0.8 (f) and 0.6 (g). The average area fraction is $\langle \phi \rangle =45 \phi_\text{c}$.
	(h) Chirality reversal frequency $f$ as a function of $W_\text{ob}/W$. \Bo{$f$ is obtained by the fast Fourier transform (FFT) of $\langle v_\text{t}(t)\rangle$}. \AS{$f$ is normalized by $f_0$ (0.24 Hz), which is the frequency of an unobstructed spontaneous flow in the channel.} The average area fractions are $\langle \phi \rangle/\phi_\text{c} =$ 15 (purple squares and lines), 45 (blue circles and lines), 60 (orange triangles and lines), 75 (red diamonds and lines).  Error bars depict the standard error from different measurements of $f$.
}
\label{Fig2}
\end{figure*}

\section{II. Active flow with a rectangular obstacle}

\subsection{Experimental system}
Our experimental system is comprised of spherical colloidal polystyrene particles  confined in a circular track (see Appendix A: Experimental Methods). The particles are energized by a uniform static (DC) electric field applied perpendicular to the bottom plate of the track. Above a critical magnitude of the field, particles begin to steadily rotate (due to the electro-hydrodynamic phenomenon of Quincke rotation) around an axis parallel to the plate and roll on the surface~\cite{quincke1896ueber,tsebers1980internal,bricard2013emergence}. Uncorrelated active diffusion of Quincke rollers transforms into a collective state with coherent polar order above a critical area fraction (i.e., rescaled number density) $\phi_c$ of the rollers. In our experiments, we find the critical area fraction $\phi_c\simeq $ 0.002, see Fig.~\ref{Fig1}(a), which is lower than the previously reported value for similar systems~\cite{bricard2013emergence} due to the smaller width for our track. In a narrow track, the area fraction  of the rollers controls the phase behavior and rollers can form stable polar states with propagating dense bands.

When a small rectangular obstacle is placed in the center of the track and partially blocks the track, the state of the system changes. We consider obstacle widths $W_\text{ob}$ smaller than the track width $W$ (i.e., $W_\text{ob}/W < 1$). Although some rollers can still pass through the gap between the obstacle and the walls of the track, other rollers instead collide with the obstacle and reverse their direction of motion (Fig.~\ref{Fig1}(b). The rollers that bounce back interact with the incoming rollers, forming high-density vortex patterns in the vicinity of the obstacle. If the obstacle is sufficiently small, the overall flow of particles maintains the initial direction. As the obstacle becomes wider, the active fluid can eventually reverse the direction of flow (Fig.~\ref{Fig1}(c), Supplementary Movie 1).

The time evolution of chirality reversal induced by the obstacle is visualized via the local area fraction of the rollers, see Fig.~\ref{Fig1}(d).
The rollers initially assemble into a counterclockwise (CCW) traveling band ($t_1$). Once the band arrives at the obstacle, rollers collide into and accumulate near the obstacle ($t_2$). The intermittent pattern of high density at the obstacle grows with time ($t_3$) and eventually reaches a critical size, when rollers start to reverse the moving direction from CCW to clockwise (CW) ($t_4$). Eventually, the rollers accumulate on the other side of the obstacle ($t_5$) and the next reversal process starts. The system is in this dynamic steady-state while the energy is supplied by the electric field.

\AS{The process of chirality reversal is accompanied by a pattern comprised of self-organized vortices as visualized in Fig.~\ref{Fig2}(a-d). Vortices are formed due to confinement with a characteristic vortex size similar to the track width $W$. 
The handedness of the vortices can be either CW or CCW as illustrated by the velocity and vorticity fields in Fig.~\ref{Fig2}(c) and Supplementary Movie 2.}
The vortex closest to the obstacle exhibits the same chirality as the overall flow.
A plausible explanation for this is due to an active analog of the centrifugal effect~\cite{bricard2015emergent}. In a polar active fluid confined in a circular track, due to the centrifugal effect, rollers near the boundary curving inwards (i.e., the outer wall of the track) continuously adjust their direction due to collisions with the outer wall, whereas rollers near a boundary with positive curvature (i.e., inner wall of the track) do not collide with the inner wall and adjust the direction only when they encounter rollers from outer layers, resulting in a radial density gradient. The presence of a radial density gradient implies that there are more rollers next to the outer gap (i.e., the gap between obstacle and outer wall of the track). If these rollers next to the outer wall bounce off the obstacle and change their direction towards the inside of the ring, a vortex is formed with the same handedness as the overall flow in the track. When the vortices are stable and localized, vortices rotating in the same direction are separated by vortices rotating in the opposite direction. The entire chain then possesses an alternating (i.e.,  antiferromagnetic) order (Fig.~\ref{Fig2}(b)). This is because on the boundary between a CW vortex and a CCW vortex, rollers move toward the same direction (Fig.~\ref{Fig2}(d), Supplementary Movie 2), thus minimizing the number of particle collisions.

With the overall flow being CW \AS{in Fig.~\ref{Fig2}(a)-(d)}, some rollers escape from the vortices and pass \Bo{around} the obstacle. Meanwhile, incoming rollers blocked by the vortices give rise to new vortices at the tail of the shock. If the number of rollers passing through the obstacle
is higher than the number of rollers being blocked, the vortex structure will dissipate over time and the overall flow will maintain the same direction (Fig.~\ref{Fig1}(b)).
By contrast, if the number of blocked rollers  is higher than the number passing through, the overall flow is being reduced by the obstacle and the number of localized vortices grows (Fig.~\ref{Fig1}(c)).
In this situation, the polar order inside the fluid weakens because the number of rollers outside the vortex region decreases. Then, the rollers in the outermost vortex escape by following the pressure gradients and the flow begins to reverse. The escaped rollers move CW, generating a local CW flow that pulls along additional rollers. Eventually, the vortex structure collapses into a CW traveling band and chirality fully reverses (see Supplementary Movie 1-2).

\AS{To determine the conditions of the chirality reversal, we vary the obstacle width  $W_\text{ob}$ and the average area fraction of rollers in the system $\langle \phi \rangle$ (see Fig.~\ref{Fig2}(e)). In general, since the high area fractions $\langle \phi \rangle$ have a stronger polarity it takes a wider obstacle to reverse the flow compared to the lower area fractions.}
Wider obstacles scatter more particles and localize more vortices. When $W_\text{ob}/W$ is large, the geometry approaches that of the split-ring resonator, and the traveling band of rollers bounces back and forth as shown in Fig.~\ref{Fig2}(f). In this case, we measure the reversal frequency to be $f$ = 0.17 Hz, which is only slightly lower than rotating frequency of polar liquid in a track without an obstacle (0.24 Hz). In this limit, the reversals are similar to those of traveling bands in a straight narrow track~\cite{bricard2013emergence}.
On the other hand, when $W_\text{ob}/W$ is small, \Bo{the obstruction effect} is too small to reverse the polarity of the active fluid.

\AS{At the phase boundary, chirality reversal occurs via a slow and complex process. While some rollers pass \Bo{around} the obstacle, others are blocked. A number of vortices are formed in front of the obstacle accumulating the large number of rollers but keeping the overall polarity of the flow. The number of vortices that a formed during the reversal process depends on the obstacle size. The dependence, however, is non monotonic: the number of vortices first increases with the obstacle size, reaches the maximum at a certain obstacle size and starts to decrease again (see Supplementary Figure 2). When obstruction becomes too large vortices disappear, and the traveling band of rollers bounces back and forth. Figure~\ref{Fig2}(g) shows plateaus of the average tangential velocity $\langle v_\text{t} \rangle$ followed by sharp reversal transitions. The polar order parameter quantified by $\langle v_\text{t} \rangle$ recovers to the same amplitude value after each reversal. In this regime, the reversal frequency $f$ is controlled by the lifetime of this unique state consisting of multiple vortices.} Thus, by tuning  $W_\text{ob}/W$ and $\langle \phi \rangle$, we can control the reversal frequency as shown in Fig.~\ref{Fig2}(h).

\subsection{Numerical model}
To show that chirality switching in annular channels is a general phenomenon inherent to polar active matter, we perform hydrodynamic simulations.
Surprisingly, the simulations reproduce the phase diagram for switching behavior despite only containing a few hydrodynamic parameters and treating the polar active fluid as a continuum. 
Our finite-element simulations track the area fraction $\phi(\mathbf{r}, t)$ and velocity $\mathbf{v}(\mathbf{r}, t)=v_0\mathbf{p}(\mathbf{r}, t)$ (where $v_0$ is the speed of a single self-propelled particle and $\mathbf{p}$ is the polarization, i.e., the local average orientation of the self-propelled particles) using a minimal version of the Toner-Tu equations~\cite{toner1995long,toner1998flocks}:
\begin{align}
  \partial_t\phi+v_0\nabla\cdot(\phi\mathbf{p})&=0  \label{equationtoner1}\\
    \partial_t\mathbf{p}+\lambda v_0(\mathbf{p}\cdot\nabla)\mathbf{p}&=(\alpha(\phi)-\beta|\mathbf{p}|^2)\mathbf{p}-\frac{v_1}{\phi_0}\nabla\phi+\nu\nabla^2\mathbf{p}
    \label{equationtoner2}\\
    \alpha(\phi)&=\alpha_0(\phi-\phi_c)
    \label{equationalpha}
\end{align}
where $\phi_0$ ($=\langle \phi \rangle$) is the average area fraction, $\alpha$ is the Landau activity coefficient favoring the spontaneous breaking of rotational symmetry due to polarization, $\beta$ is a nonlinear drag that stabilizes a \Bo{statistically} steady state with finite polarization, and $\nu$ is the effective viscosity. The speed $v_1$ controls the speed of sound $c$ in the polar active fluid, $c = \sqrt{v_0 v_1}$. The presence of the coefficient $\lambda$ is a consequence of the breaking of Galilean invariance in the active fluid, with the case $\lambda = 1$ corresponding to the restored Galilean invariance in the convection term of the Navier-Stokes equations. \Anton{Without Galilean invariance, other convective terms are possible, including the term $\lambda_2 v_0 \nabla |\mathbf{p}|^2/2$~\cite{marchetti2013hydrodynamics}. The addition of such terms does not change the qualitative conclusions of our simulations.} Equation~(\ref{equationtoner1}) is the continuity equation that stems from the conservation of particle number. Equation~(\ref{equationtoner2}) is the equation of motion, which is solved for the flow profile. 
\Anton{Note that Refs.~\cite{toner1995long, toner1998flocks} introduce a pressure in Eq.~(\ref{equationtoner2}) and an additional equation of state relating pressure and density. We instead assume that (as in an ideal gas) the pressure is proportional to the density and eliminate the pressure from the equations of motion, see e.g. Ref.~\cite{geyer2018sounds}. }
Equation~(\ref{equationalpha}) corresponds to the equation of state for the activity: the activity coefficient $\alpha$ changes sign when the area fraction $\phi$ crosses the critical area fraction $\phi_c$. At low densities, $\alpha$ is negative and the polarization tends to zero, whereas at high densities, $\alpha$ is positive and we observe flocking behavior.

We use the continuum hydrodynamic approach due to two of its advantages: (1) The computational time for the finite-element simulations depends only weakly on the particle area fraction $\phi(\mathbf{r}, t)$, unlike a particle-based model. This allows us to efficiently simulate cases corresponding to large particle numbers. (2) The hydrodynamic model provides a minimal description and contains only a few parameters. 
\Anton{From the remarkable qualitative agreement between simulations and experiment, we therefore conclude that these phenomena do not require fine tuning of experimental and theoretical parameters. }
In simulations, we keep constant the parameters $\alpha_0$ ($=5$), $\phi_c$ ($=0.4$), $\beta$ ($=1$), $\nu$ ($=1.7$), $v_0$ ($=1$), $v_1$ ($=10\phi_0$), and $\lambda$ ($=0.8$).
\Anton{These parameters are chosen so that Eqs.~(\ref{equationtoner2}--\ref{equationalpha}) qualitatively describe a polar active fluid. In some cases, these choices set a convenient scale for simulations (e.g., $\beta$ and $v_0$). In other cases, we choose the parameters to be of the same order as existing experimental measurements, for example the values of $\nu$, $\lambda$, and the speed of sound, which have been measured in Ref.~\cite{geyer2018sounds}. Although the experimental setup of the Quincke-roller fluid in Ref.~\cite{geyer2018sounds} differs from ours, we believe that these parameters should roughly be of the same order in our experiments.}
We explore the parameter space by changing the average area fraction $\langle \phi \rangle$ (relative to critical area fraction $\phi_c$) and the geometry of confinement.

\begin{figure*}[t]
\centering
\includegraphics[width=1.0\linewidth]{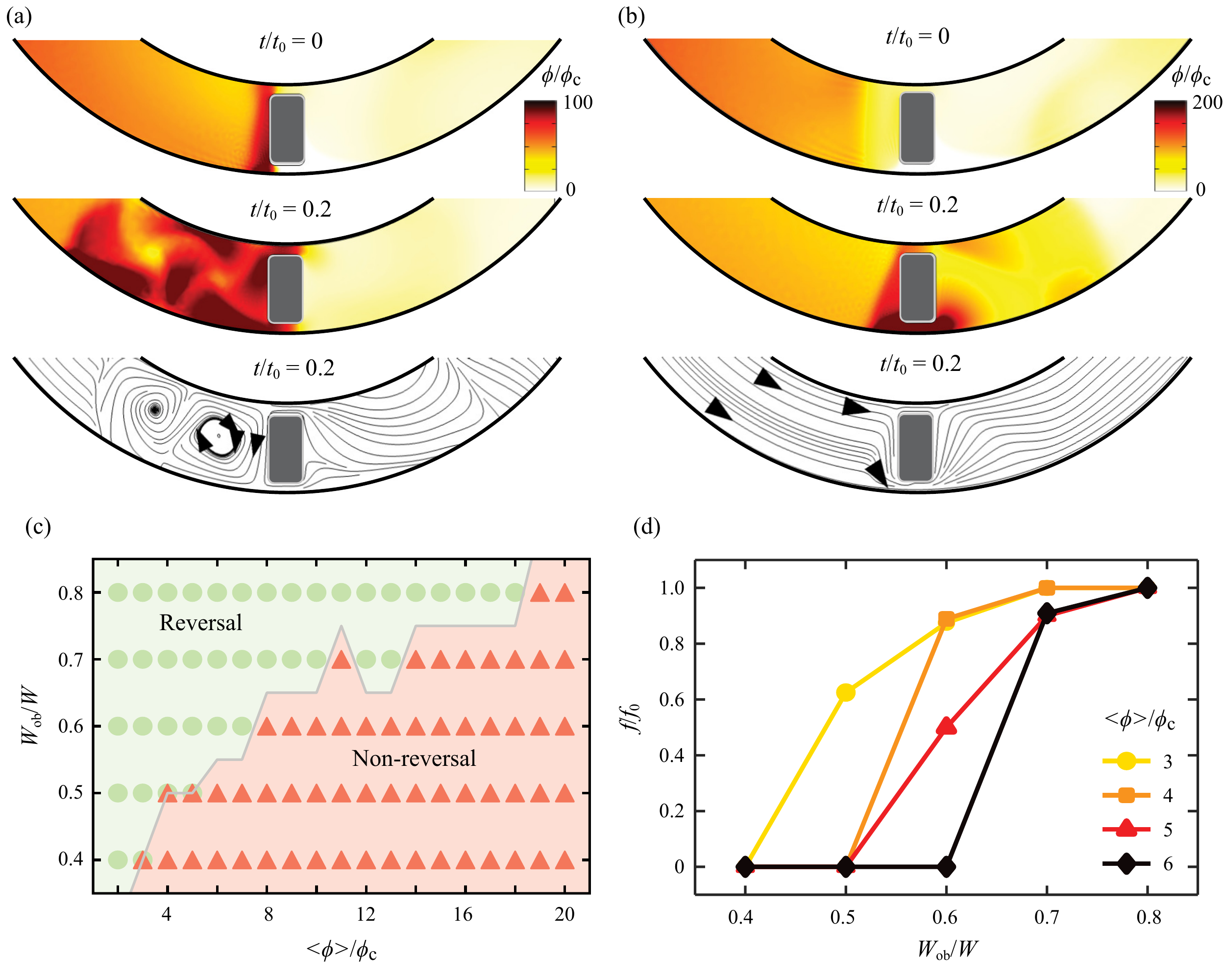}
\caption{
Simulations for chirality switching.
	(a) Top: At $t = 0$, active fluid moves counterclockwise and collides with a rectangular obstacle (obstacle width $W = 0.8 W_0$, where $W_0$ is channel width).
Middle: At $t = 0.2 t_0$ (where $t_0$ is the total simulation run time) and at low to medium densities [here, the average area fraction of the fluid
$\langle \phi \rangle = 10 \phi_\text{c}$, where $\phi_\text{c}$ is the critical area fraction in Eq.~(\ref{equationalpha})], this front bounces off the obstacle and switches direction, traveling away clockwise. Bottom: Contour plot at $t = 0.2 t_0$ shows the characteristic shock-reversal pattern consisting of short-lived vortices.
	(b) Same as (a), but at higher densities ($\langle \phi \rangle = 20 \phi_\text{c}$). Top: At $t = 0$, an active matter front moves counterclockwise towards the obstacle (again, $W = 0.8 W_0$). Middle: At $t = 0.2 t_0$ and at high densities, the front passes around the obstacle and continues to move counterclockwise, leaving a jammed states around the obstacle. Bottom: Contour plot at $t = 0.2 t_0$ shows the flow around the obstacle as the front is passing through.
	(c) Phase diagram showing whether the system exhibits chirality reversal as a function of obstacle width (normalized by channel width) and average area fraction (normalized by critical area fraction). In agreement with experiments, reversal behavior occurs at large obstacle widths and low densities.
	(d) To quantify the transition from non-reversal to reversal, the frequency $f$ of reversal is plotted against rectangle width for four values of average area fraction ($\langle \phi \rangle/\phi_\text{c} = $ 3, 4, 5, and 6). The frequencies are normalized by the frequency $f_0 = f(W_\text{ob}/W = 0.8)$ (i.e., the right-most point). The plot shows a sharp transition at an area-fraction-dependent obstacle width.
}
\label{Fig3}
\end{figure*}

To make parallels with the experiment, we use the same geometry of annular confinement.
In the case of the rectangular obstacle (Fig.~\ref{Fig3}), we use the same values of the proportion $W_{ob}/W$ of obstacle width $W_{ob}$ to channel width $W = 0.25 R_{o}$, where $R_{o}$ is the outer radius of the channel.
We keep the length of the obstacle constant at a value of $0.1 R_{o}$.
We avoid hydrodynamic singularities at the obstacle corners by rounding (i.e., filleting) them with a small radius of curvature $0.0125 R_{o}$.
The initial area fraction is set to zero near the obstacles to make no-slip boundary conditions easy to implement in the entire geometry \Anton{independently of obstacle size and shape}.
The initial velocity profile is set tangential to the ring (and counterclockwise).
Because we search for \Bo{statistically} steady states, we find the \Anton{qualitative conclusions that we draw from the results of our} simulations to be insensitive to details of the initial conditions. As the relaxation time depends on the other simulation parameters, we also vary the total simulation run time $t_0$.
Throughout our simulations, we only change the geometric parameter $W_{ob}/W$, the average area fraction $\langle \phi \rangle/\phi_c$, and the simulation run time $t_0$.


Figure~\ref{Fig3} shows the results from simulations with a rectangular obstacle. The simulations allow us to visualize impacts of the polar active fluid onto an obstacle with a good degree of spatial resolution, see Figure~\ref{Fig3}a.
Although some of the active fluid manages to bypass the obstacle, most of the matter reverses direction. To better interpret the flow pattern during the reversal, the bottom snapshot (Fig.~\ref{Fig3}(a)) shows \AS{streamlines for the same point in time, where each line traces the flow path. From this plot, we observe that the shock front interacts with the bulk of the fluid to create vortices familiar from the experiments}. These vortices are unstable and decay to a laminar flow profile shortly after the shock front leaves the region (see Supplementary Movie 3).

 Figure~\ref{Fig3}(b) shows an equivalent selection of snapshots to Fig.~\ref{Fig3}(a), but for the case where the chirality does not reverse. The front approaches the obstacle (top image), \Bo{gets obstructed} slightly at the obstacle (middle image), but ultimately does not reverse, as can be observed from the streamline plot at the bottom. An interesting characteristic that can only be seen from the bottom two plots in Fig.~\ref{Fig3}(b) is that the majority of the fluid passes through the \Bo{outer gap} with little fluid passing through the \Bo{inner} gap.
 This effect can be interpreted as a type of centrifugal effect created as a result of simulating a polar active fluid obeying Eqs.~(\ref{equationtoner1}--\ref{equationalpha}) and following a circular path. Note that this centrifugal effect arises despite the overdamped dynamics of the fluid.
 \Anton{An intuitive explanation for this analogy is that the left-most term in Eq. ~(\ref{equationtoner2}), $\partial_t\mathbf{p}$ mimics the acceleration term in an inertial system because $\mathbf{p} = \mathbf{v}$.}
 This observation suggests an origin for the clockwise chirality of the vortex neighboring the obstacle in Fig.~\ref{Fig3}(a).
 \Anton{We note that in contrast with experiments, the vortex nearest the obstacle in simulations has a chirality opposite to the overall flow inside the track. We hypothesize this difference arises because in experiments, this vortex chirality is controlled by the density gradient, whereas in simulations chirality is controlled by velocity gradients. In turn, these two contributions lead to opposite for vortex chirality.}
 
 The phase diagram for simulated active fluids shown in Fig.~\ref{Fig3}(c) bears striking similarity to the experimental phase diagram in Fig.~\ref{Fig2}(e).
 Switching tends to occur at higher obstacle widths and at lower densities.
 The mechanism for suppressing chirality switching at higher densities is the one
 shown in Fig.~\ref{Fig3}(a-b). At higher densities, the higher polarization
 of the impacting front helps the fluid tunnel through the narrow channels around the obstacle.
 The area-fraction dependence of polarization is a general feature of polar active fluids described by the Toner-Tu equations~(Eqs.\ref{equationtoner1}--\ref{equationalpha}), which is why this hydrodynamic description captures the qualitative features of chirality switching without any input on the microscopic details of the experimental system.

 In addition to general trends, the phase diagram Fig.~\ref{Fig3}(c)
 shows several distinctive features of these simulations.
 One such feature is the spike in the boundary within Fig.~\ref{Fig3}(c) corresponding to the values $\langle \phi \rangle =11 \phi_\text{c}$ and $W_\text{ob}=0.7 W$.
 At this point, which is colored in red, the chirality does not switch; however, at higher area fraction, the systems reenters the green chirality-switching region.
 The mechanism for this reentrant behavior is apparent by observing the simulation dynamics (see Supplementary Movie 3):
 the spike corresponds to a resonant condition for the active fluid.
 The resonant condition can be understood by comparing the time that the part of the fluid stuck at the obstacle takes to reverse and the time the rest of fluid front takes to travel around the ring.
 In the Supplementary Movie 3, we observe that for the point ($\langle \phi \rangle =11 \phi_\text{c}$, $W_\text{ob}=0.7 W$),
 part of the front escapes through the bottleneck, travels around the ring, and impacts the obstacle a second time just as the remaining fluid reverses. This second impact of the traveling front then pushes more of the active fluid through the bottleneck. 

 A different type of switching transition can occur at low densities, which we designate as hybrid points through which we draw the transition line in Fig.~\ref{Fig3}(c).
 At these hybrid points, the chirality of the flow is undefined for much of the simulation run time. This is because the amount of fluid that leaks \Bo{around} the obstacle and keeps its chirality is comparable to the amount of fluid that reverses flow and switches chirality.
 Therefore, the simulations show two fronts moving in opposite directions (see Supplementary Movie 3).
 We contrast these hybrid cases with the low-frequency cases for the transition boundary at higher densities (and obstacle widths). For the low-frequency cases, the front goes around the ring several times before a distinct chirality-switching event occurs. Because there is only one moving front, the chirality of the flow is well defined throughout.
 The transition boundary in Fig.~\ref{Fig3}(c) is drawn such that it is equidistant between the well-defined reversal and non-reversal points, but passes through the hybrid points.

 To quantify the simulations from Fig.~\ref{Fig3}(c), we plot in Fig.~\ref{Fig3}(d) (analogously to Fig.~\ref{Fig2}(h)) the frequency of reversal against the width of the obstacle for four different values of average area fraction. These frequencies corresponds to peaks in the Fourier transform of the simulation data for the tangential velocity as a function of time. A frequency of zero indicates no reversals, with frequency values normalized by the frequency of reversal at $W_\text{ob}=0.8 W$ (i.e., the right-most points). Similarly to Fig.~\ref{Fig2}(h) for experiment, the simulation data in Fig.~\ref{Fig3}(d) shows the general trend that the frequency of reversal increases with obstacle size and decreases with area fraction.

\begin{figure*}
\centering
\includegraphics[width=1.0\linewidth]{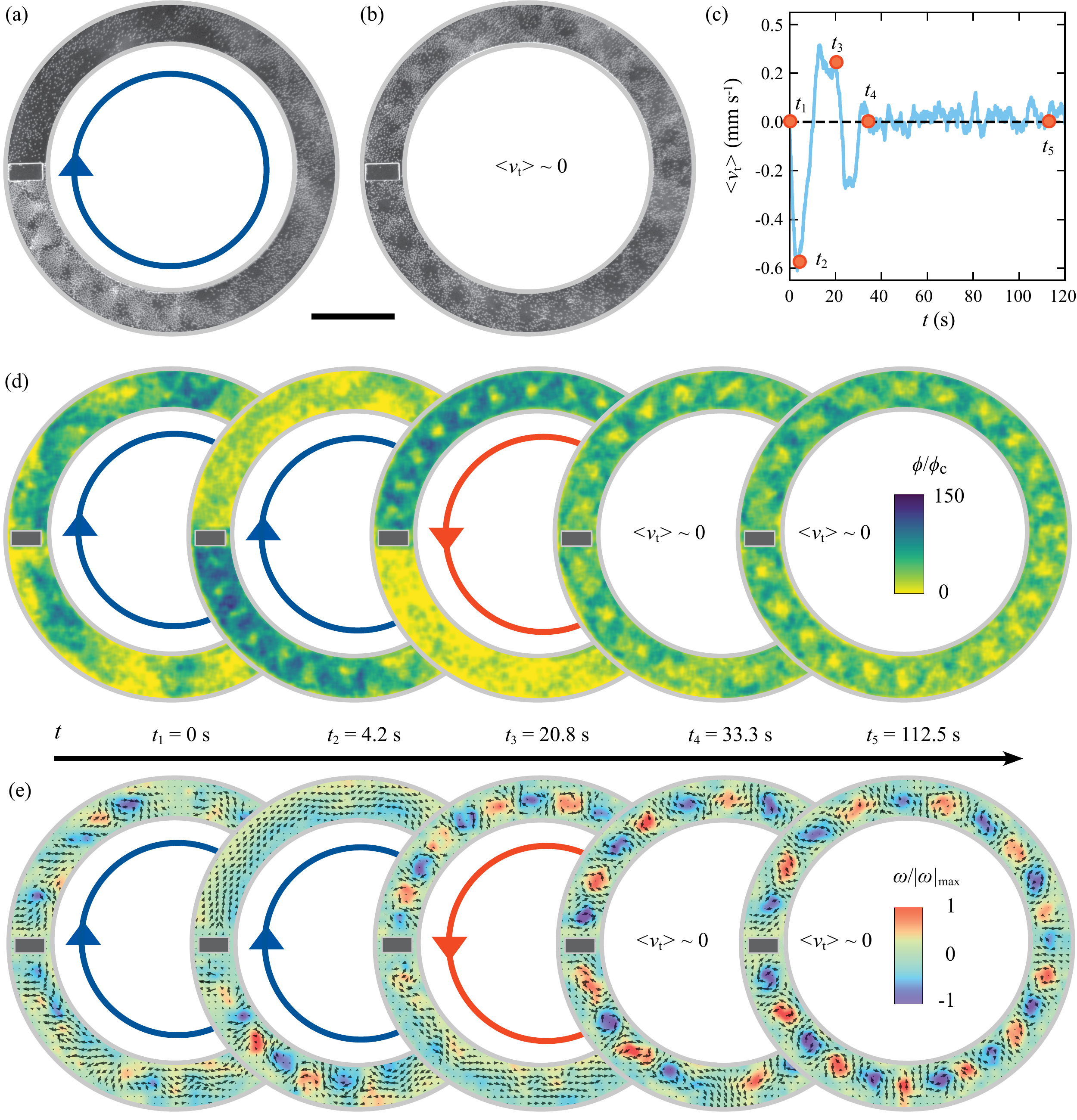}
\caption{
Apolar state of rollers.
	(a) Initial microscopic image of polar active flow in a CW direction with multiple vortices forming.
	(b) Microscopic image of \Bo{statistically} steady state: the vortices are evenly distributed and the state is apolar, i.e., $|\langle v_\text{t} \rangle|\sim$ 0. See also Supplementary Movie 4.
	(c) Time evolution of average tangential velocity $\langle v_\text{t}  \rangle$ during the formation of the apolar vortex state. The system is polar when $|\langle v_\text{t}  \rangle|>$ 0 and reaches an apolar state when $|\langle v_\text{t} \rangle|\sim$ 0. Five red dots indicates the positions of 5 different stages shown in (d-e).
	(d) Corresponding evolution of local particle densities.
	(e) Corresponding evolution of velocity (arrows) and vorticity (background colors) fields.
	In (c-e), rollers start to move ($t_1$); form a CW polar state ($t_2$); reverse to be in a CCW polar state ($t_3$); form multiple vortices fully distributed throughout the track ($t_4$); and finally relax to multiple vortices with an alternating order ($t_5$).
	Blue and red circles with an arrow show the CW and CCW flows, respectively. The parameter values are $W_\text{ob}$ = 0.8 $W$ and $\langle \phi \rangle =90 \phi_\text{c}$. The scale bar is 0.5 mm.
}
\label{Fig4}
\end{figure*}

\section{III. Apolar state with vortices}

In experiments, a unique apolar state appears when both obstacle width $W_\text{ob}/W$ and area fraction $\langle \phi \rangle$ are large (Fig.~\ref{Fig2}(e), Fig.~\ref{Fig4}, and Supplementary Movie 4). After the DC electric field is applied, rollers first assembly into a CW polar liquid. When the polar liquid collides with the obstacle, multiple vortices are formed in a mechanism similar to the transient state formed during a reversal process (Fig.~\ref{Fig4}(a), $t_2$ in Fig.~\ref{Fig4}(d-e)).
Neighboring vortices counter-rotate, so that overall this state does not exhibit any chirality nor an overall polar order.

In a \Bo{statistically} steady-state reversal process described in the previous section, chirality changes sign while the strength of polar order (i.e., $|\langle v_\text{t} \rangle|$) is the same before and after a reversal event.
By contrast, on approach to the apolar vortex state, the global polar order diminishes after each collision with a wide obstacle. As the area occupied by localized vortices grows, this loss of global polar order accelerates. To understand this mechanism, consider a single collision event. Similar to the case in which the fluid reverses its \Bo{statistically} steady-state chirality, the polar liquid switches the flow direction from CW to CCW (e.g., snapshot $t_3$ in Fig.~\ref{Fig4}(d-e)).
However, unlike the situation in a \Bo{statistically} steady-state reversal process, the strength of polar order in the fluid ($|\langle v_\text{t} \rangle|$) decreases during each reversal event, as shown in Fig.~\ref{Fig4}(c).  The polar order parameter ($|\langle v_\text{t} \rangle|$) eventually relaxes to 0. After a few reversals, the region occupied by the vortices grows and instead of a localized set of vortices next to the obstacle, the vortex state permeates throughout most of the ring. When only a small space remains for the traveling band, it quickly bounces around and assembles into vortices ($t_4$ in Fig.~\ref{Fig4}(d-e)). In the \Bo{statistically} steady-state, vortices occupy the whole track and destroy any oscillations of polar order (Fig.~\ref{Fig4}(b), $t_5$ in Fig.~\ref{Fig4}(d-e)).

This apolar state observed in experiments does not appear in the hydrodynamic simulations of the Toner-Tu model. Instead in the simulations, even when a large number of vortices is formed, eventually this vortex region near the obstacle destabilizes and a produces a counter-propagating front. This suggests a particle-scale origin for the stability of a vortex, not captured within the lowest-order hydrodynamic model of the Toner-Tu equations~(\ref{equationtoner1}-\ref{equationalpha}). For example, such quantities as the number of particles per vortex could play a crucial role in vortex stability but cannot be tracked in the continuum.
Nevertheless, the simulations do produce vortices, which suggests that the vortex-based mechanism of reversals is general beyond the specific experimental realization that we consider.


\section{IV. Triangular obstacles}
\begin{figure*}
\centering
\includegraphics[width=1.0\linewidth]{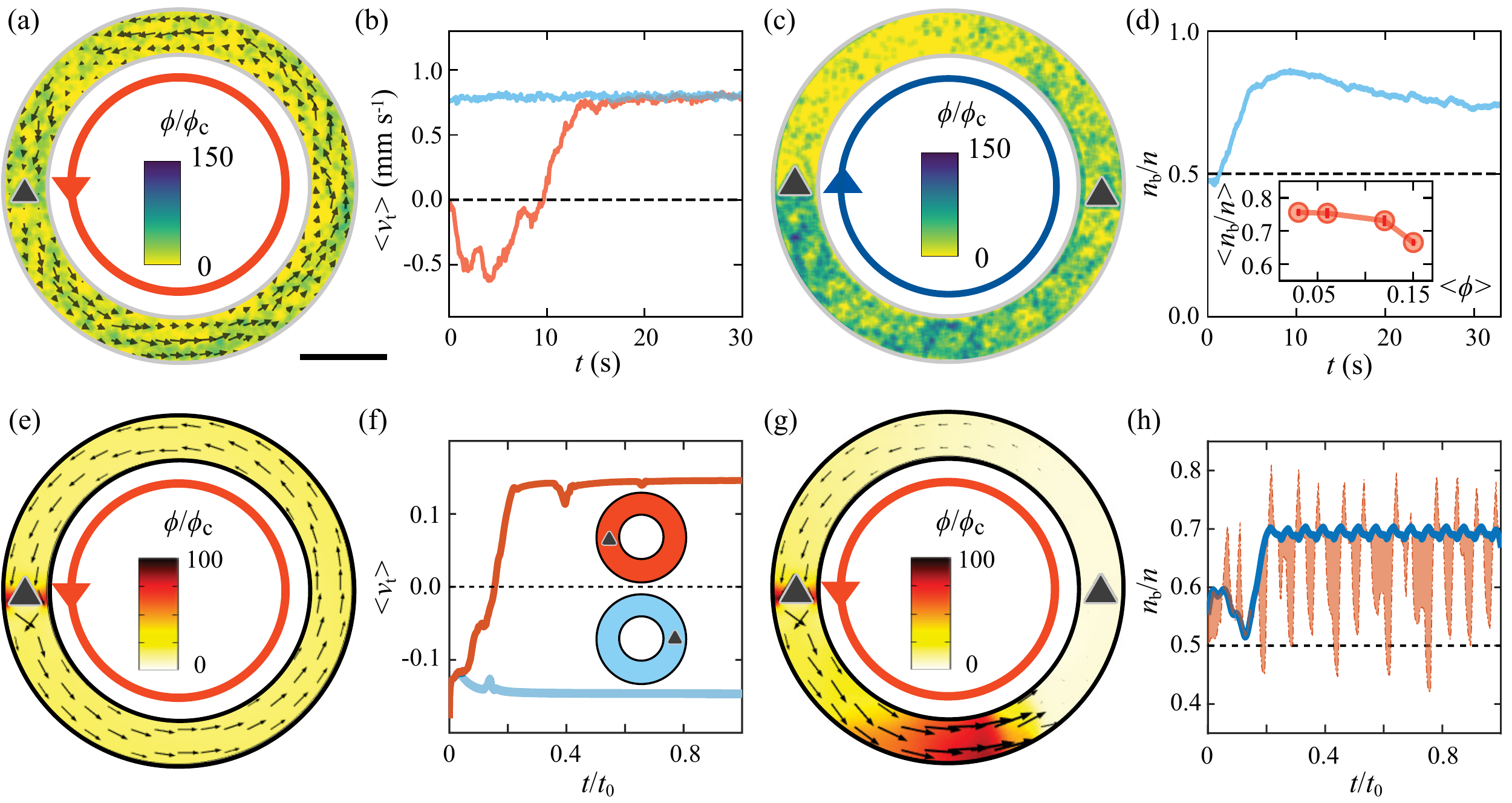}
\caption{
Chirality reversal induced by asymmetric obstacles.
	(a) Expertimental results for velocity field (arrows) and local particle area fraction (background color) of rollers in a track with a triangular obstacle pointing up. In the \Bo{statistically} steady state, particles move CCW (red circle with an arrow) due to the asymmetric obstacle. The parameters are $W_\text{ob}$ = 0.6 $W$ and $\langle \phi \rangle =30 \phi_\text{c}$. The scale bar is 0.5 mm. See also Supplementary Movies 5.
	(b) Time evolution of average tangential velocities $\langle v_\text{t} \rangle$ of rollers in a track shown in (a) starting from either a CW (red) or a CCW (blue) flow. In the \Bo{statistically} steady state for both initial conditions, $\langle v_\text{t} \rangle$ eventually flows in a CCW direction.
	(c) Local area fraction of rollers in a track with two triangular obstacles both pointing up. Particles accumulate in the bottom half of the track. The parameters are $W_\text{ob}$ = 0.8 $W$, $\langle \phi \rangle =30 \phi_\text{c}$, and experimental run time is $t_0$ = 30 s. See also Supplementary Movie 5.
	(d) The percentage of particles $n_\text{b}/n$ in the bottom half of the track using data from experiments shown in part (c). The quantities $n$ and $n_\text{b}$ are the total number of particles in the track and the number of particles in the bottom half of the track, respectively. In this geometry, $n_\text{b} > n$. The insert shows $\langle n_\text{b}/n \rangle$ at different area fractions $\langle \phi \rangle$. Error bars are calculated from the standard deviation of $n_\text{b}/n$ once $n_\text{b}/n$ reaches the \Bo{statistically} steady-state plateau.
	(e) Snapshot of simulated active fluid with triangular obstacle on the left side of the ring. Velocity arrows show direction of flow and color scale represents the local area fraction of active particles. $\langle \phi \rangle =10 \phi_\text{c}$.
	(f) Time evolution of fluid velocities $v_\text{t}$ tangential to the track in terms of total simulation run time~$t_0$. The initial conditions are the same in both geometries, with initial flow counterclockwise, but the triangle was placed on either the left (red) or right (blue) side of the ring. In the \Bo{statistically} steady state, the fluid flows from triangle tip to base in both cases.
	(g) Snapshot of simulated active liquid with two triangular obstacles on both the left and right sides of the ring. Because of the flow pattern around the triangles, the integrated area fraction $n = \int \phi d {\bf x}$ is higher on the bottom of the ring, as shown in (h) using the ratio $n_\text{b}/n$ of the integrated area fraction in the bottom half of the ring as a proportion of the total. In (h), a moving average (blue line) is taken over the area fraction fluctuations (shaded orange area). $\langle \phi \rangle =10 \phi_\text{c}$.
}
\label{Fig5}
\end{figure*}

Rectangular obstacles allow us to quantify how spontaneous chiral order becomes destroyed, but these obstacles maintain the overall achiral geometry of the container. Specifically, the ring with a rectangular obstacle favors neither CW nor CCW flow, which might not be true for inclusions having a generic shape. In order to quantify how chiral geometries affect the symmetry of flow in the \Bo{statistically} steady state, we instead consider inclusions shaped like an equilateral triangle (see Fig.~\ref{Fig5}).

Experimentally, we placed a triangular obstacle so that the CCW flow encounters a pointy tip (vertex), whereas the CW flow is hindered by the base (i.e., side of the triangle, see Fig.~\ref{Fig5}(a)). Within a large region of parameter values (such as triangle size and roller area fraction), the \Bo{statistically} steady-state flow direction is always such that the fluid flows from triangle tip to triangle base, i.e., CCW in this experimental set up (Fig.~\ref{Fig5}(b), Supplementary Movie 5). In the case of rectangular obstacles, the initial conditions are what determines the chirality of \Bo{statistically} steady-state flow. By contrast, in the case of flow around a triangular obstacle, chirality is determined by the geometry and is independent of initial conditions, as shown by the time evolution of $\langle v_\text{t} \rangle$, i.e., the average velocity in the ring, in Fig.~\ref{Fig5}(b) and Supplementary Movie 5. This explicit breaking of chiral symmetry then allows for the design of active-fluid-based microfluidic devices based on unidirectional flow and transport.

Asymmetric obstacles can also be used to concentrate active particles based on confinement geometry alone. This coupling between geometry and flow allows us to design low-density and high-density regions that are nevertheless able to exchange particles, a property that distinguishes active fluids from their equilibrium counterparts. Fig.~\ref{Fig5}(c) shows the local area fraction of rollers in a track with two up-pointing triangular obstacles. Rollers are concentrated in the bottom half of the track because the bases of the triangles reverse particles and prevent them from passing  with only  a small amount of particles leaking to the top half of the track. The time evolution of this accumulation process is shown in Fig.~\ref{Fig5}(d) and Supplementary Movie 5. Initially, rollers are evenly distributed throughout the track. When energized, the escape rate of rollers from the top half to the bottom half is much higher than the reverse rate, resulting in rollers accumulating in the bottom half of the ring. In the \Bo{statistically} steady state, the rate for these two flows balances out and the portion of particles in the bottom half, $n_\text{b}$, reaches a plateau (with approximately 75\% of particles in the bottom half of the ring, i.e., $n_\text{b}/n \sim 0.75$). Note that rollers still move collectively and oscillate in the half track (see Supplementary Movie 5).  As the average area fraction of rollers  increases in the track, the  effects of having finite-sized rollers come into play (for example, excluded volume interactions and jamming), and the effect of obstacles on particle concentration becomes weaker, see Fig.~\ref{Fig5}(d) insert.


The simulations largely reproduce these experimental results and allow us to generalize these conclusions to any polar active fluid. The snapshot in Fig.~\ref{Fig5}(e) shows the case in which an asymmetric obstacle is placed inside the simulation channel such that the initial front approaches from the top and moves from triangle tip to triangle base (see Supplementary Movie 6 for dynamics). This snapshot is taken after the \Bo{statistically} steady state has been established and both the area fraction and velocity fields no longer change in time. The simulation shows that in this case, the obstacle does not reverse the chirality of fluid flow. Instead, the obstacle provides a bottleneck, which can be seen from the increased area fraction around the corners at the triangle's base.

Figure~\ref{Fig5}(f) represents the time evolution for tangential velocities in two cases: the triangle positioned on the left and on the right part of the track. The initial conditions of the fluid are identical for both geometries: the fluid starts with the same velocity and area fraction profiles, having counterclockwise initial flow. The blue line (corresponding to triangle located on the left) shows the situation captured by the snapshot in Fig.~\ref{Fig5}(e): the chirality never reverses and the system finds a \Bo{statistically} steady state with the same counterclockwise flow as the initial conditions. The orange line (corresponding to triangle located on the right) shows a reversal event early in the simulation: the tangential velocity changes from negative to positive. The resulting \Bo{statistically} steady state has the opposite chirality from the initial condition, with the final flow following the direction from triangle tip to triangle base (see Supplementary Movie 6 for dynamics). These simulations demonstrate that, in general, asymmetric obstacles can be used to control the chirality of active-fluid flow in a ring.

Fig.~\ref{Fig5}(g-h) illustrate corresponding simulation results for an active fluid confined in a ring with two asymmetric obstacles. Similar to the experimental observations, more of the fluid is trapped at the bottom half of the ring between the bases of the triangles (see Supplementary Movie 6 for dynamics).
The plot in Fig.~\ref{Fig5}(h) quantifies this effect by showing $n_\text{b}/n$, the integrated area fraction in the bottom half of the ring as a portion of the total, as a function time. The value for $n_\text{b}$ is calculated via $n_\text{b} = \int \phi({\bf x}) d{\bf x}$ by integrating over the bottom half.
The simulation data is shown as the dashed line, strongly fluctuating when the fronts impact the triangular obstacles. The moving average of the data is plotted as the blue line, which quickly relaxes to a constant value of $n_\text{b}/n \approx 0.7$, a value consistent with experiments.
\Anton{We expect the value of $n_\text{b}/n$ to depend on obstacle shape. For example, if obstacle size is decreased so that the flow does not reverse when impacting the obstacle, then the fluid area fraction would remain evenly distributed between the two halves.}

\section{V. Conclusion}

In summary, we have used a combination of experiments and simulations  to investigate and  quantify  collective chiral states of a confined polar active fluid with scatterers. We have found that chiral states are stable in the presence of small symmetric obstacles. However, large obstacles can destabilize a chiral state even if they do not explicitly break chiral symmetry. For example, rectangular obstacles lead to \Bo{statistically} steady states in which the handedness of fluid flow oscillates between clockwise and counterclockwise. We also show that the frequency of chirality reversals can be tuned by the size of the obstacle and the area fraction of active particles. Asymmetric obstacles exemplified by triangular scatterers favor one chirality and can enforce a \Bo{statistically} steady-state whose chirality is independent of initial conditions. We have examined emergent apolar \Bo{statistically} steady states in rings with obstacles, from an array of counter-rotating vortices to uneven dynamic density distributions caused by a barrier composed of two triangles. Our simulations of Toner-Tu hydrodynamics not only confirm experimental results, but also suggest that these phenomena are independent of the microscopic details of the active constituents and, therefore, generalize to a broad range of polar active fluids.

Chiral symmetry often plays a defining role in the mechanics of active materials~\cite{bricard2013emergence, kaiser2017flocking, driscoll2017unstable, kokot2017active, soni2019odd, martinez2018emergent, kokot2018manipulation, banerjee2017odd}. Our finding demonstrate how to translate confinement geometry into a chiral motion.
Instead of considering scatterers that destroy coherent motion, we show how control over scatterer configuration can generate coherence in flow patterns. These results pave the way towards controlling transport and collective behavior in active systems that live in complex environments. Our detailed experimental and numerical characterization can be directly translated into the design of dynamic chiral components for  active materials and devices. Such devices based on active fluids can range from microfluidic mechanical components exhibiting topologically protected sound waves~\cite{souslov2017topological,souslov2019topological} to logic gates based on channels filled with active fluids~\cite{woodhouse2017active}. Taken together our work provides new insights into the collective behavior and control of active colloidal materials in complex environments, and demonstrates how passive scatterers can be used to manipulate dynamic patterns and active polar states  in active fluids in general.

\section{Acknowledgments}

\begin{acknowledgments}
The research of Alexey Snezhko and Bo Zhang was supported by the U.S. Department of Energy, Office of Science, Basic Energy Sciences, Materials Sciences and Engineering Division. Use of the Center for Nanoscale Materials, an Office of Science user facility, was supported by the U.S. Department of Energy, Office of Science, Office of Basic Energy Sciences, under Contract No. DE-AC02-06CH11357.
This collaboration was initiated at the Aspen Center for Physics, which is supported by National Science Foundation grant PHY-1607611.  The participation of Anton Souslov at the Aspen Center for Physics was supported by the Simons Foundation. Anton Souslov acknowledges the support of the Engineering and Physical Sciences Research Council (EPSRC) through New Investigator Award No.~EP/T000961/1.
\end{acknowledgments}

\section{Appendix A: Methods}

\subsection{1. Experimental setup and procedures}

Polystyrene colloidal particles (G0500, Thermo Scientific) with an average diameter of 4.8 \SI{}{\micro\meter} are dispersed in a 0.15 mol L$^{-1}$ AOT/hexadecane solution. The colloidal suspension is then injected into a track constructed by SU-8 blocks and two parallel ITO-coated glass slides (IT100, Nanocs). The thickness of the SU-8 blocks is 45 \SI{}{\micro\meter}. The outer diameter $D$ and width $W$ of the track are 2 mm and 0.25 mm, respectively. A rectangular or equilateral triangular SU-8 block with same height is placed in the center of the track to be an obstacle. The rectangular obstacle has a fixed length, $L$, of 0.1 mm and a width, $W_{ob}$, varying from 0.1 mm to 0.2 mm. The triangular obstacle has a size of 0.15 mm or 0.2 mm.

A uniform DC electric field is applied to the cylinder chamber through two ITO-coated glass slides.  The electric field is supplied by a function generator (Agilent 33210A, Agilent Technologies) and a power amplifier (BOP 1000M, Kepco Inc.). The field strength $E$ is kept at 2.7 V \SI{}{\micro\meter}$^{-1}$. The sample cell is observed under a microscope with a 4$\times$ microscope objective \Bo{or a 2$\times$ microscope objective with a 1.6$\times$ magnifier} and videos are recorded by a fast-speed camera (IL 5, Fastec Imaging). The frame rates are 120 frames per second (FPS) for particle image velocimetry (PIV) and 851 FPS for particle tracking velocimetry (PTV). PIV, PTV and further data analysis are carried out with custom codes in Python as well as Trackpy \cite{allan2016trackpy}. \Bo{Typically, the velocity fields are calculated by PIV except Fig.~\ref{Fig2}(c). The PIV sub-windows are 24 $\times$ 24 pixels (83 $\times$ 83 \SI{}{\micro\meter}) or 48 $\times$ 48 pixels (166 $\times$ 166 \SI{}{\micro\meter})  and have 50\% overlap. The velocity field in Fig.~\ref{Fig2}(c) is calculated by the average of actual particle velocity from PTV in each sub-window. The sub-window is 32 $\times$ 32 pixels (40 $\times$ 40 \SI{}{\micro\meter}) and have zero overlap.
The reversal frequency, $f$, is obtained by the fast Fourier transform (FFT) of $\langle v_\text{t}(t)\rangle$. A typical spectrum usually has a well-defined central peak that is taken as a characteristic $f$ at a specific experimental condition. The peak is sharp for fast reversals and becomes broader for slow reversals near the phase boundary. See Supplementary Figure S1 demonstrating typical spectra.}

%


\end{document}